\documentclass[conference,
]{IEEEtran}

\usepackage{amsmath,amssymb,amsfonts}
\usepackage{algorithmic}
\usepackage{graphicx}
\usepackage{textcomp}
\usepackage{xcolor}

\usepackage{dcolumn}
\usepackage{bm}
\usepackage{upgreek}
\usepackage[sort&compress, numbers]{natbib}
\begin{document}

\title{Numerical study of SQUID array responses due to asymmetric junction parameters}



\author{\IEEEauthorblockN{1\textsuperscript{st} M. A. Gal\'i Labarias}
\IEEEauthorblockA{ \textit{CSIRO Manufacturing}\\
 Lindfield, NSW, Australia. \\
marc.galilabarias@csiro.au}
\and
\IEEEauthorblockN{2\textsuperscript{nd} E. E. Mitchell}
\IEEEauthorblockA{ \textit{CSIRO Manufacturing}\\
 Lindfield, NSW, Australia. \\
 emma.mitchell@csiro.au}
}

\maketitle

\maketitle

\begin{abstract}
Superconducting quantum interference device arrays have been extensively studied for their high magnetic field sensitivity.
The performance of these devices strongly depends on the characteristic parameters of their Josephson junctions, i.e. their critical currents and shunt resistances.
Using a resistively shunted junction model and including thermal noise, we perform a numerical investigation of the effects of asymmetric Josephson junctions by independently studying variations in the critical currents and junction resistances.
We compare the voltage response of a dc-SQUID with a 1D parallel SQUID array and study the maximum transfer function dependence on the number of junctions in parallel, the screening parameter and thermal noise strength.
Our results show that the maximum transfer function and linearity increase with the number of junctions in parallel for arrays with different junction resistances, in contrast to SQUID arrays with identical junctions or with spreads in the critical currents.

\end{abstract}


%
\IEEEpeerreviewmaketitle

\section{Introduction}

Superconducting quantum interference devices (SQUIDs) are highly sensitive magnetic field sensors which rely on their superconducting properties and the Josephson effect.
Understanding the effect of each of the components of the SQUID, such as inductance, normal resistance and the effects of thermal noise are necessary to design effective devices.
For a dc-SQUID these have been extensively studied \cite{Tesche1977, Clarke2004}. 
However SQUID arrays present a more complex system to study due the larger number of variables such as the number of junctions \cite{Oppenlander2000, Gali2022a} and the superconducting film properties, such as the London penetration depth \cite{Tolpygo1996, Keenan2021}. 
At the same time, this large parameter space allows for more tuneability and has the potential to further increase their performance.
The effect of the junction characteristics, i.e. critical current and resistance spreads, can substantially affect the device performance.
Spreads in the junction parameters can appear due to fabrication processes and they are very common in high-temperature superconductors \cite{Shadrin2003, Lam2014}. For these junction spreads the junction critical current and normal resistance have been shown to be correlated \cite{Gross1997, Yoshida2004}.
Alternatively, asymmetric junctions and shunt resistances can be deliberately designed in order to obtain a voltage response with high linearity \cite{Jeng2005, Kornev2009a, Drung2009, Crete2021}. 

In this paper we numerically investigate the voltage response and maximum transfer function of SQUID arrays with non-identical Josephson junctions using a resistively shunted junction model \cite{Gali2022a}.
Here we consider the junction asymmetry due to different critical currents and different shunt resistances independently, similar to the study done by \citet{Tesche1977} for a dc-SQUID.
This method allows us to discern between the asymmetric contributions of critical currents and shunt resistances which is important when designing high performance SQUID arrays.

\section{Modelling}
In this paper we use a resistively shunted model \cite{Gali2022a} to perform a numerical investigation of SQUID arrays with non-identical Josephson junctions that are uniformly biased (one bias lead aligned with each junction).
The expression for the dynamics of the array is defined as, 
\begin{align}
    \overrightarrow{\frac{d \varphi}{d \tau}} = \widehat{r}\left[\vec{i}_n - \widehat{i_c} \; \overrightarrow{\sin (\varphi)} + \frac{\Phi_0}{2\pi I_c}  \widehat{K} \widehat{L}^{-1} \widehat{D}\vec{\varphi} + \vec{C} \right] \; ,
    \label{eq:ODE}
\end{align}{}
where $\vec{\varphi}=(\varphi_1, \varphi_2, \dots, \varphi_{N_p})$ with $\varphi_k$ the phase difference at the k$^{th}$ junction, $\tau$ is the normalized time, $\Phi_0$ the flux quantum and $\vec{C} = \left( \widehat{K}_I^{-1}\vec{I}_b - \widehat{K} \widehat{L}^{-1} \vec{\Phi}_{nf} \right)/I_c$ is a vector with time-independent components. $\widehat{r} (j,k)=r_k \, \delta_{j,k}$ where $r_k=(R_k/R)$ is the normalized resistance of the k$^{th}$ junction and $R$ is the average resistance across the array. $\widehat{i_c} (j,k) = i_{c_{k}}\,  \delta_{j,k}$ with $i_{c_{k}}= I_{c_{k}} / I_c$ the normalized critical current of the k$^{th}$ junction and and $I_c$ is the average critical current of the array.
The definition for the matrices $\widehat{K}, \; \widehat{L},\; \widehat{D}$, $\widehat{K}_I$ and vector quantities $\vec{\Phi}_{nf}$ and $\vec{I_b}$ can be found in \cite{Gali2022a}.

The normalized noise currents are generated at each time-step using random number generators that follow a Gaussian distribution where its mean and mean-square-deviation satisfy
\begin{align}
    \overline{i_{n,k}} = 0  \qquad  \text{ and } \qquad  \overline{i^2_{n,k}} = 2\Gamma_k / \Delta \tau \; , 
\end{align}

where the thermal noise strength $\Gamma_k$ is
\begin{equation}
    \Gamma_k= \frac{1}{r_k}\frac{2\pi k_B T}{ \Phi_0 I_c} \;.
    \label{eq:Gamma}
\end{equation}

Here $k_B$ is the Boltzmann constant, $T$ the device operating temperature and $\Delta \tau$ the normalized time-step used when solving Eq. (\ref{eq:ODE}) numerically. 
In this work we use $\Delta \tau = 0.1$.

The junction asymmetry can be caused by different critical currents or by having junctions with different shunt resistances.
To analyse independently the contributions of the critical currents and resistances we define $i_{c_k}$ and $r_k$ using two independent parameters $\alpha$ and $\rho$, which define the degree of asymmetry of the junction.

Here $\alpha$ defines the normalized degree of asymmetry of the junction's critical currents. For this study, the junction critical currents are defined by 
\begin{equation}
    i_{c_k} = 1 - \alpha + \frac{2 \alpha}{N_p - 1}(k-1),
\end{equation}
where $N_p$ is the number of junctions in parallel.
Likewise the junction shunt resistances are defined using a normalized degree of asymmetry $\rho$,
\begin{equation}
    \frac{1}{r_k} = 1 - \rho + \frac{2 \rho}{N_p - 1}(k-1).
\end{equation}

\subsection{Normalized variables}
The normalized bias current is $i_b=I_b/(N_p I_c)$ where $I_b$ is the total bias current. 
In this work we are studying uniformly biased SQUID arrays, where each bias lead carries $I_b/N_p$.
The normalized applied flux is $\phi_a= \Phi_a/\Phi_0$ with $\Phi_a$ being the applied magnetic flux, and the $RI_c$ normalized time-averaged voltage is $\bar{v}$. 
The screening parameter is $\beta_L=2L_s I_c/\Phi_0$ and the thermal noise strength $\Gamma = \langle \Gamma_k \rangle= 2\pi k_B T/(\Phi_0 I_c)$ is the average of $\Gamma_k$.

\section{Results}

Here we will study the contribution of these asymmetries independently by fixing the degree of asymmetry of the critical currents to $\alpha=0$ while varying the resistances degree of asymmetry $\rho$, and vice-versa.


\subsection{Voltage response}
\begin{figure}
    \centering
    \includegraphics[width=0.5\textwidth]{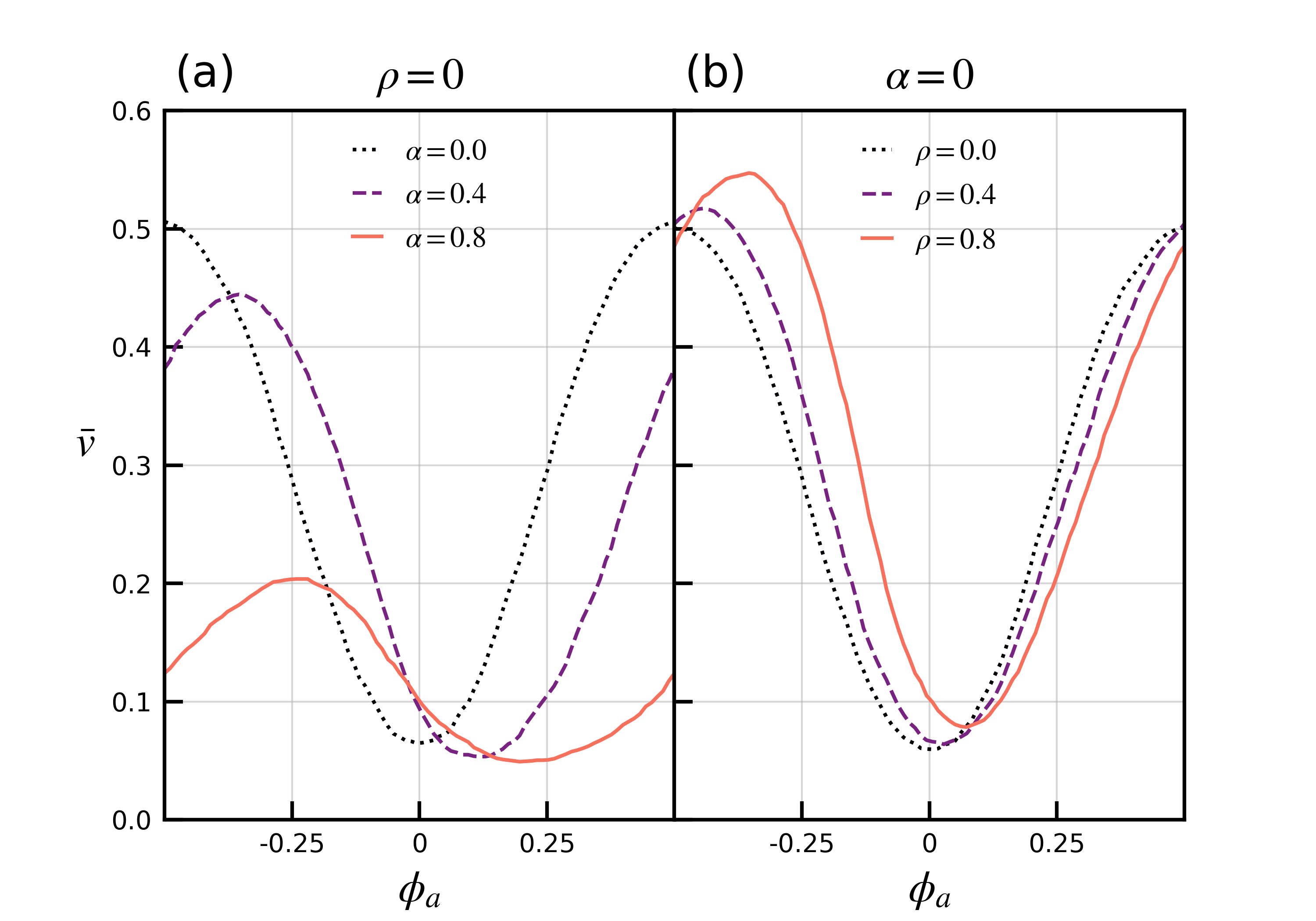}
    \caption{Normalized voltage versus normalized applied flux of a dc-SQUID with asymmetric junctions. (a) Asymmetric critical currents defined by $\alpha=0$ (black dotted line), $\alpha=0.4$ (purple dashed line) and $\alpha=0.8$ (orange line) with equal resistances $\rho = 0$. (b) Asymmetric shunt resistances defined by $\rho=0$ (black dotted line), $\rho=0.4$ (purple dashed line) and $\rho=0.8$ (orange line) with equal critical currents $\alpha=0$. Here $\beta_L=0.75$, $\Gamma=0.16$ and $i_b=0.75$.}
    \label{fig:v-phi_Ns1Np2}
\end{figure}

Figure \ref{fig:v-phi_Ns1Np2} shows the $\bar{v}-\phi_a$ response of a dc-SQUID with different $\alpha$ and $\rho$ for a device with $\beta_L=0.75$, $\Gamma=0.16$ and $i_b=0.75$.
Figure \ref{fig:v-phi_Ns1Np2}(a) shows the effect of varying the critical current degree of asymmetry while fixing $\rho=0$. Increasing $\alpha$ produces a decrease of the voltage modulation depth and a $\phi_a$-shift.
Figure \ref{fig:v-phi_Ns1Np2}(b) shows the effect of varying $\rho$ while keeping $\alpha=0$. In this case increasing $\rho$ increases the voltage response asymmetry and also produces a smaller $\phi_a$-shift.
These results are consistent with previously shown results for a dc-SQUID without thermal noise effects \cite{Tesche1977}.

\begin{figure}[h!]
    \centering
    \includegraphics[width=0.5\textwidth]{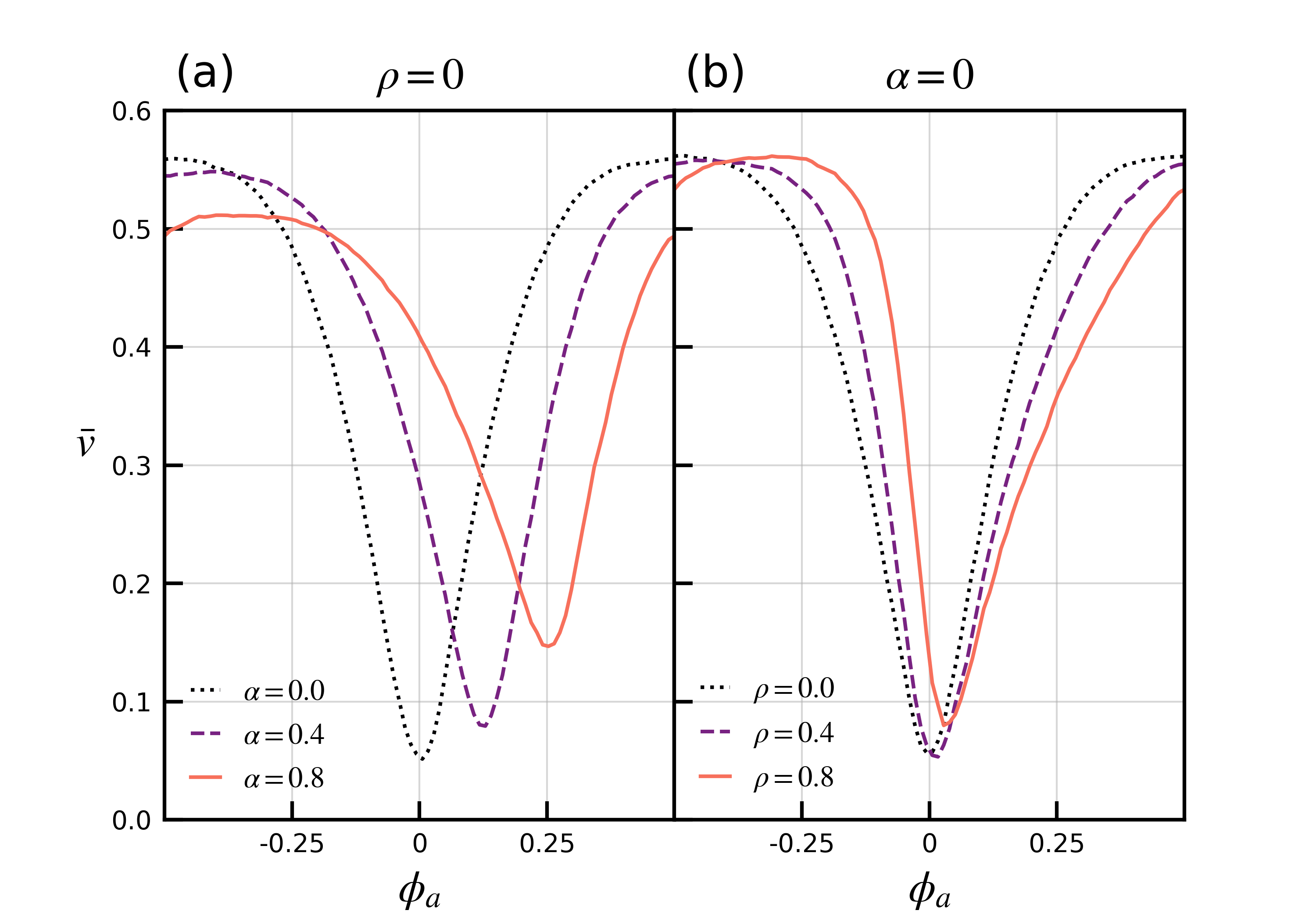}
    \caption{Normalized voltage versus normalized applied flux of a 1D parallel SQUID array with $N_p=10$ and asymmetric junctions. (a) Asymmetric critical currents defined by $\alpha=0$ (black dotted line), $\alpha=0.4$ (purple dashed line) and $\alpha=0.8$ (orange line) with equal resistances $\rho = 0$. (b) Asymmetric shunt resistances defined by $\rho=0$ (black dotted line), $\rho=0.4$ (purple dashed line) and $\rho=0.8$ (orange line) with equal critical currents $\alpha=0$. Here $\beta_L=0.75$, $\Gamma=0.16$ and $i_b=0.75$.}
    \label{fig:v-phi_Ns1Np10}
\end{figure}

Figure \ref{fig:v-phi_Ns1Np10} shows the $\bar{v}-\phi_a$ response of a 1D parallel SQUID array with $N_p=10$ with asymmetric junctions, $\beta_L=0.75$, $\Gamma=0.16$ and $i_b=0.75$.
Figure \ref{fig:v-phi_Ns1Np10}(a) shows the effect of varying the critical current degree of asymmetry $\alpha$ while fixing $\rho=0$. Increasing $\alpha$ produces a decrease of the voltage modulation depth and a $\phi_a$-shift in the voltage response. 

Figure \ref{fig:v-phi_Ns1Np10}(b) shows the effect of varying $\rho$ while keeping $\alpha=0$. Similar to the dc-SQUID case, increasing $\rho$ 
creates a more asymmetric voltage response.

Comparing Fig. \ref{fig:v-phi_Ns1Np2} and Fig. \ref{fig:v-phi_Ns1Np10}, the voltage modulation depth reduction with $\alpha$ is less pronounced for the $N_p=10$ SQUID array, while the same $\phi_a$-shift is produced. The level of asymmetry and linearity of the voltage response due to $\rho$ is larger for the SQUID array than for the dc-SQUID.

\subsection{Maximum transfer function dependence on $N_p$}

\begin{figure}
    \centering
    \includegraphics[width=0.5\textwidth]{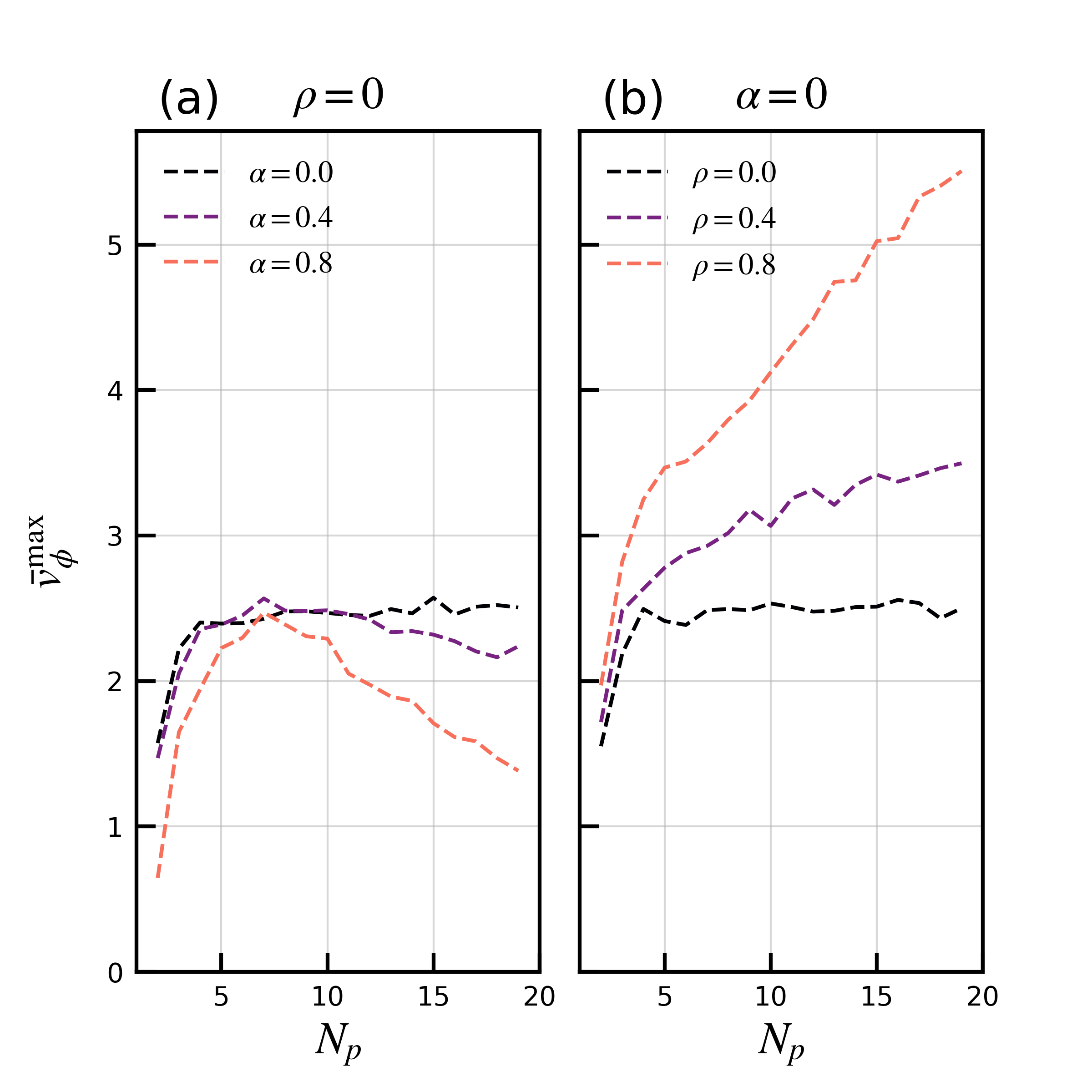}
    \caption{Maximum transfer function versus number of junctions in parallel. (a) Asymmetric critical currents defined by $\alpha=0$ (black dotted line), $\alpha=0.4$ (purple dashed line) and $\alpha=0.8$ (orange line) with equal resistances $\rho = 0$. (b) Asymmetric shunt resistances defined by $\rho=0$ (black dotted line), $\rho=0.4$ (purple dashed line) and $\rho=0.8$ (orange line) with equal critical currents $\alpha=0$. Here $\beta_L=0.75$, $\Gamma=0.16$ and $i_b=0.75$.}
    \label{fig:dvdp-vs-Np}
\end{figure}

Figures \ref{fig:v-phi_Ns1Np2} and \ref{fig:v-phi_Ns1Np10} showed that the voltage modulation depth and voltage response asymmetry increases with $N_p$. 
To better understand this dependence Fig. \ref{fig:dvdp-vs-Np} shows the maximum transfer function $\bar{v}_{\phi}^{\max}=\max_{\phi_a} \left[ \partial \bar{v}/\partial \phi_a \right]$ versus $N_p$ for devices with $\beta_L=0.75$, $\Gamma=0.16$ and $i_b=0.75$.
Figure \ref{fig:dvdp-vs-Np}(a) shows $\bar{v}_{\phi}^{\max}-N_p$ curves for $\alpha=0, \; 0.4$ and $0.8$ and $\rho=0$. For $N_p>10$, $\bar{v}_{\phi}^{\max}$ decreases with $N_p$ for $\alpha>0$. The decrease of $\bar{v}_{\phi}^{\max}$ with $N_p$ is stronger for larger $\alpha$.

Figure \ref{fig:dvdp-vs-Np}(b) shows $\bar{v}_{\phi}^{\max}-N_p$ curves for $\rho=0, \; 0.4$ and $0.8$ and $\alpha=0$. 
In contrast to Fig. \ref{fig:dvdp-vs-Np}(a), increasing $\rho$ allows $\bar{v}_{\phi}^{\max}$ to increase with $N_p$ past the plateauing point present for identical junctions, i.e. $\alpha=\rho=0$ \cite{Gali2022b}.
This is a key result of our paper, which shows that the performance of a parallel SQUID array uniformly biased can be further enhanced by introducing junction asymmetry due to different shunt resistances.

\subsection{Maximum transfer function dependence on $\beta_L$ and $\Gamma$}
In previous sections $\beta_L$ and $\Gamma$ have been fixed to typical values of YBCO SQUIDs with step-edge junctions operating at 77 $K$.
In this section the degree of asymmetry parameters $\alpha$ and $\rho$ are fixed and the dependence of the maximum transfer on $\beta_L$ and $\Gamma$ is studied.

\begin{figure}
    \centering
    \includegraphics[width=0.5\textwidth]{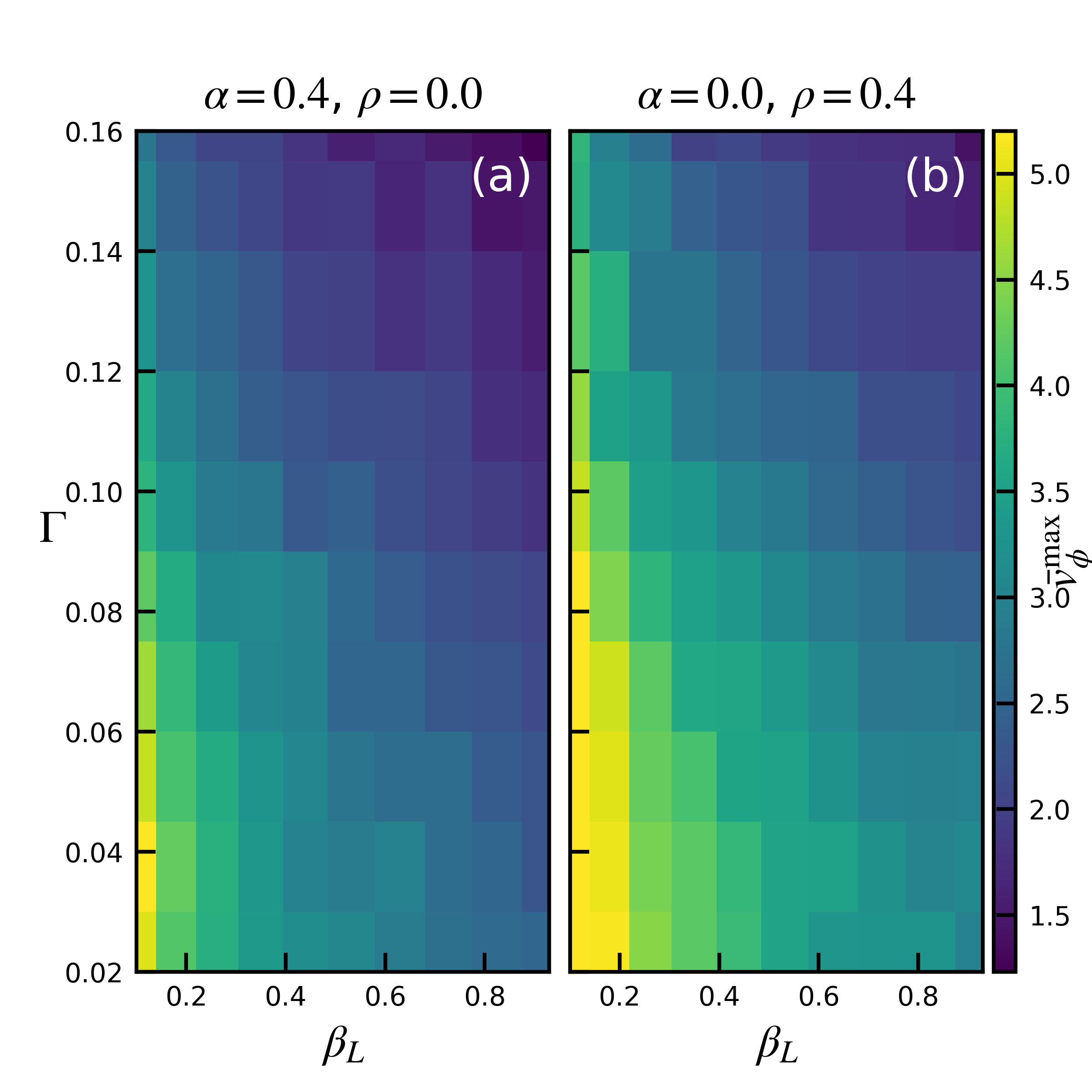}
    \caption{3D map of the maximum transfer function versus $\beta_L$ and $\Gamma$ of a dc-SQUID junction asymmetry due to (a) different critical currents: $\alpha=0.4$ and $\rho=0$ and (b) different shunt resistances : $\alpha=0$ and $\rho=0.4$.}
    \label{fig:colormap_dcSQUID}
\end{figure}

Figure \ref{fig:colormap_dcSQUID} shows a 3D map of the $\bar{v}_{\phi}^{\max}$ of a dc-SQUID for a range of $\beta_L=0.1, \dots, 0.9$ and $\Gamma=0.02, \dots, 0.16$.
The dependence of $\bar{v}_{\phi}^{\max}$ on $\beta_L$ and $\Gamma$ is very similar for a device with critical current asymmetry (Fig.\ref{fig:colormap_dcSQUID}(a)) and with shunt resistance asymmetry (Fig. \ref{fig:colormap_dcSQUID}(b)), with the larger $\bar{v}_{\phi}^{\max}$ appearing at small $\beta_L$ and $\Gamma$.

\begin{figure}
    \centering
    \includegraphics[width=0.5\textwidth]{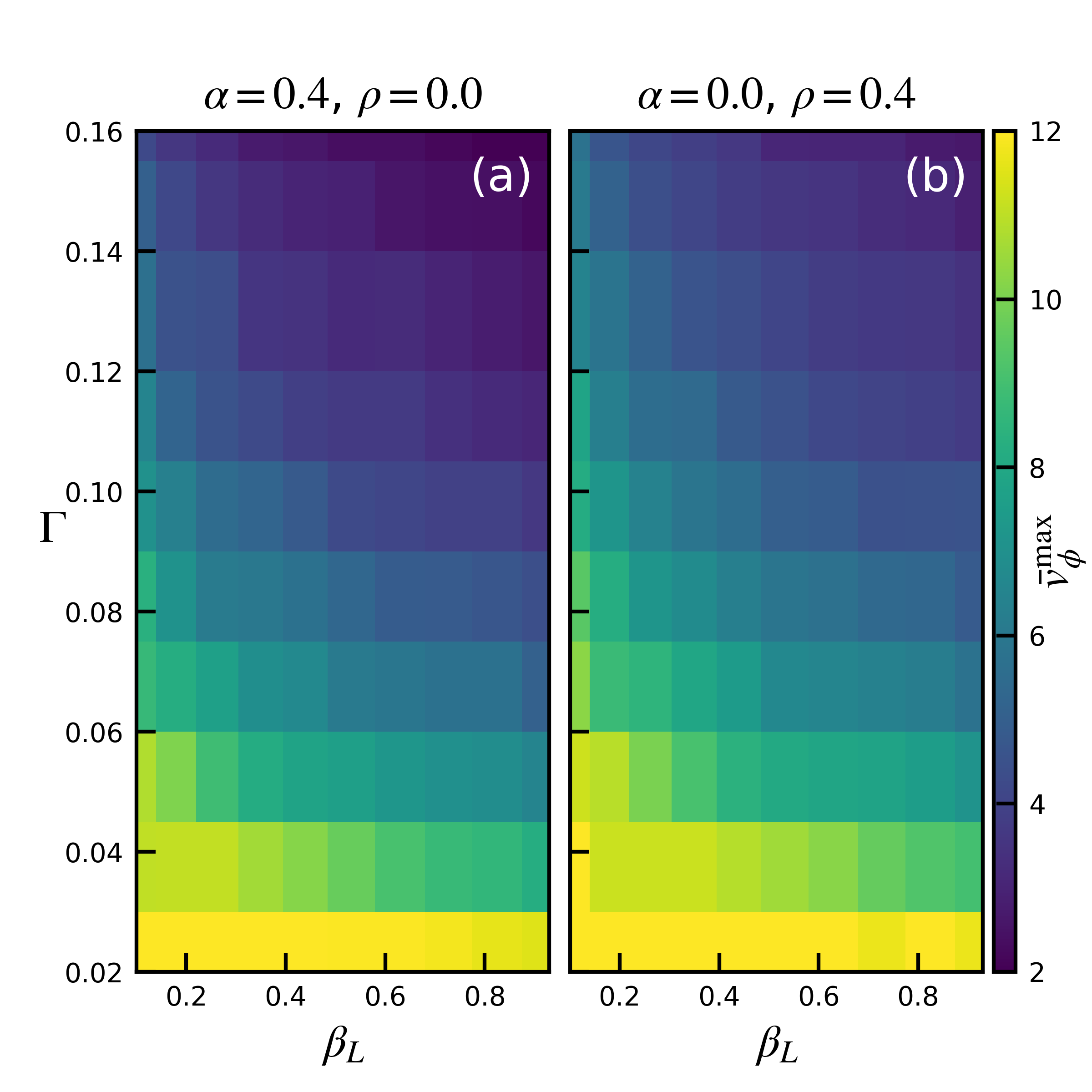}
    \caption{3D map of the maximum transfer function versus $\beta_L$ and $\Gamma$ of a 1D parallel SQUID array with $N_p=10$ and junction asymmetry due to (a) different critical currents: $\alpha=0.4$ and $\rho=0$ and (b) different shunt resistances: $\alpha=0$ and $\rho=0.4$.}
    \label{fig:colormap_SQUIDarray}
\end{figure}

Figure \ref{fig:colormap_SQUIDarray} shows a 3D map of the $\bar{v}_{\phi}^{\max}$ of a 1D parallel SQUID array with $N_p=10$ for a range of $\beta_L=0.1, \dots, 0.9$ and $\Gamma=0.02, \dots, 0.16$.
As seen before for the dc-SQUID, the $\beta_L$ and $\Gamma$ dependence of Fig. \ref{fig:colormap_SQUIDarray}(a) and Fig. \ref{fig:colormap_SQUIDarray}(b) is similar.
However, for the 1D parallel array the most relevant parameter is $\Gamma$ showing that large transfer functions can be obtained for different $\beta_L$ if $\Gamma$ is kept small.

\section{Summary}
In this paper we have discussed the effects of junction asymmetries caused by different critical currents and different shunt resistances on the response of 1D SQUID arrays.

Firstly, we have compared the voltage versus magnetic flux response of a dc-SQUID with a 1D parallel SQUID array with $N_p=10$ and showed that asymmetric junctions reduce the voltage modulation depth and produce a shift with the applied flux in the voltage response.
Our simulations also showed that asymmetric junctions due to different resistances produce a voltage response asymmetry, which becomes more pronounced for SQUID arrays.

Secondly, we analysed the maximum transfer function as a function of $N_p$. Our results showed that, after reaching a maximum, $\bar{v}_{\phi}^{\max}$ starts to decrease with $N_p$ for arrays where their junctions have different critical currents.
On the other hand, when the junction asymmetry is due to different shunt resistances, but equal critical currents $\bar{v}_{\phi}^{\max}$ keeps increasing with $N_p$, over the studied range. 
Therefore having different shunt resistances could overcome the plateauing trend present for SQUID arrays with identical junctions.

Finally, we investigated the $\bar{v}_{\phi}^{\max}$ dependence on $\beta_L$ and $\Gamma$. Our simulations showed that a dc-SQUID requires small $\beta_L$ and $\Gamma$ to optimise the transfer function, while SQUID arrays show more robustness for larger $\beta_L$ if $\Gamma$ is kept small.

\section*{Acknowledgement}
The authors are grateful to C. Lewis, K-H. M\"uller and J. Beyer for helpful discussions.




\ifCLASSOPTIONcaptionsoff
  \newpage
\fi




\bibliography{non-identical-JJ}

%








\end{document}